\documentclass[12pt]{JHEP3}
\usepackage{amsmath,epsfig,graphicx}

\newcommand{\be}{\begin{equation}}
\newcommand{\ee}{\end{equation}}
\newcommand{\bq}{\begin{eqnarray}}
\newcommand{\eq}{\end{eqnarray}}
\newcommand{\bsq}{\begin{subequations}}
\newcommand{\esq}{\end{subequations}}
\newcommand{\bc}{\begin{center}}
\newcommand{\ec}{\end{center}}
\newcommand{\pd}{\partial}

\title{Cosmic String Dynamics and Evolution in Warped Spacetime}

\author{A. Avgoustidis \\ 
{\it Departament ECM and ICCUB, Universitat de Barcelona, Diagonal 647, 
08028 Barcelona, Spain}\\ 
{\quad\quad\quad\quad\quad\quad\quad\quad and} \\  
{\it DAMTP, CMS, University of Cambridge,\\
Wilberforce Road, Cambridge CB3 0WA, United Kingdom\\}
\email{tasos@ecm.ub.es}}

\abstract{
We study the dynamics and evolution of Nambu-Goto strings in a 
warped spacetime, where the warp factor is a function of the internal 
coordinates giving rise to a `throat' region.   The microscopic equations 
of motion for strings in this background include potential and 
friction terms, which attract the strings towards the bottom of the 
warping throat.  However, by considering the resulting macroscopic 
equations for the velocities of strings in the vicinity of the throat, 
we note the absence of enough classical damping to guarantee that the 
strings actually reach the warped minimum and stabilise there.  Instead,  
our classical analysis supports a picture in which the strings experience 
mere deflections and bounces around the tip, rather than strongly damped 
oscillations.  Indeed, 4D Hubble friction is inefficient in the internal 
dimensions and there is no other classical mechanism known, which could
provide efficient damping.  These results have potentially important 
implications for the intercommuting probabilities of cosmic superstrings.   
}

\keywords{cosmic strings, string cosmology}
\preprint{UB-ECM-PF-07-35\\ DAMTP-2007-121}

\begin{document}

 \section{Introduction}

  Recent progress in studying inflation in string theory and 
  supergravity has led to a notable revival of interest in 
  cosmic strings \cite{Kibblerev,PolchIntro,DavKib}.  It is 
  now believed that cosmic string formation is generic in both 
  supersymmetric grand unified theories \cite{Jeannerot} and  
  brane inflation \cite{SarTye,DvalVil,PolchStab}, where the 
  inflaton potential is of the hybrid type \cite{hybrid} and 
  the inflationary phase ends with a phase transition, leaving 
  behind a network of topological (or semilocal \cite{UrrAchDav}) 
  cosmic strings.  Although such string networks cannot be 
  solely responsible for the formation of the observed structure 
  in the universe \cite{Battye,Battye1}, they can still act as 
  subdominant contributors.  Indeed, a recent study~\cite{BHindKU} 
  finds that a $\Lambda$CDM model with a flat spectrum of scalar 
  perturbations and a network of (field theory simulated) cosmic 
  strings contributing to the Cosmic Microwave Background (CMB) 
  anisotropy at the $10\%$ level, provides an excellent fit to 
  the observational data (see also Refs.~\cite{Battye,Battye1, 
  Battye2,Battye3} for related work).  On the other hand, 
  significantly higher contribution from strings is inconsistent 
  with the CMB, so, thinking on the positive side, one can use 
  this fact to constrain the parameter space of inflationary 
  models with network production.  Perhaps more optimistically, 
  one can hope to identify characteristic observational 
  signatures of the string networks appearing in different 
  models, and then try to look for them observationally, 
  in an attempt to point out a direction towards the correct 
  class of models.      

  From this point of view, brane inflation is of particular interest,
  being one of the most well-developed inflationary models in string 
  theory, and also producing cosmic strings with distinct properties.   
  Indeed, string networks in this setup evolve in a higher-dimensional 
  space, and this can have important effects on the probability of 
  string intercommutation \cite{JoStoTye2,PolchProb,Jackson} and 
  on string velocities \cite{EDVOS}, resulting in a significant 
  enhancement of network string densities \cite{intprob,Sak}.  
  Further, the networks produced in these scenarios are of the 
  $(p,q)$-type \cite{DvalVil,PolchStab}, and strings can interact 
  in more complicated ways than ordinary Abelian cosmic strings, 
  forming $Y$-shaped junctions.  In the most well-developed models 
  \cite{KKLMMT}, the strings are evolving in a warped spacetime 
  with one or more `throat' regions, resulting from a combination 
  of $D$-branes and fluxes present in the compactification.  This 
  warping gives rise to potentials for the string positions in 
  the internal dimensions, so one expects that the strings get 
  confined in a region around the bottom of the throat. However, 
  this localisation process has not yet been studied in detail.  
  The purpose of this paper is to study the dynamics and evolution 
  of strings in the vicinity of such throat regions.  What we find 
  is that, although there are, indeed, potential terms pulling the 
  strings towards the bottom of the throat, the classical evolution 
  does not have enough (Hubble) friction to guarantee that they 
  actually fall on it and stabilise there.  Instead, depending on 
  the impact parameter and velocity, a string can simply deflect, 
  bounce and escape to infinity, enter a series of bounces, or 
  form a bound orbit around the minimum.  These possibilities can 
  have important implications for the evolution of string networks, 
  because the probability of string intercommutings is inversely 
  proportional to the effective volume available to the 
  string~\cite{PolchProb,Jackson}.  Thus, if the  strings are 
  not confined at the bottom of the warping throat, this 
  probability gets further suppressed leading to further  
  enhancement in the network density.  

  The structure of the paper is as follows.  In section \ref{dynamics}, 
  we consider the dynamics of strings evolving in a spacetime that 
  is a warped product of a Friedmann-Lema\^{i}tre-Robertson-Walker 
  (FLRW) universe with a static toroidal space.  We write down the 
  Nambu-Goto equations of motion in this background and identify a 
  number of potential and friction terms, which tend to pull the 
  strings towards highly warped regions.  In section \ref{VOS}, 
  we use these equations to develop a model for studying a simple 
  string configuration (one in which the strings are straight in 
  the internal dimensions) moving near a minimum of the warping 
  potential (section \ref{VOS}).  We point out the weakness of 
  Hubble friction and, choosing a simple warping function, we 
  solve the model for different initial conditions obtaining 
  a sample of string trajectories, which include deflections, 
  bounces and bound orbits.  In section \ref{IIB} we try to make 
  contact with more realistic IIB compactifications, by considering 
  a slightly different setup, in which the metric is a warped product  
  of Minkowski spacetime and an unspecified 6-dimensional Riemannian 
  manifold.  We perform a qualitative analysis in terms of one-dimensional 
  motion in an effective potential, finding the same general types of 
  orbits.  We discuss the appearance of Hubble friction in this 
  setup, in terms of cosmological expansion in the effective 4D 
  theory, and comment on how the results of section \ref{VOS} can 
  be understood in this picture also.  We summarise our results 
  and discuss their implications for string evolution in section 
  \ref{discuss}.

 \section{\label{dynamics}String Dynamics in Warped Spacetime} 

   We start by considering a cosmic string propagating in a warped 
   $(D+1)$-dimensional FLRW spacetime with metric 
   \be\label{warped}
    ds^2=g_{\mu\nu}{\rm d}x^\mu {\rm d}x^\nu = h^{-1/2}({\bf l}) 
    \left[N(t)^2 {\rm d}t^2-a(t)^2 {\rm d}{\bf x}^2\right]-h^{1/2} 
    ({\bf l})b(t)^2 {\rm d}{\bf l}^2 ,  
   \ee 
   where the warp factor $h$ is a function of the internal coordinates 
   ${\bf l}$.  The motion of the string generates a two-dimensional 
   timelike surface, the string worldsheet $x^\mu=x^\mu(\zeta^\alpha)$,
   parametrised by the worldsheet coordinates $\zeta^\alpha$, 
   $\alpha=0,1$.   The dynamics is given by the Nambu-Goto action
   \be\label{nambu}
    S=-\mu \! \int \! \sqrt{-\gamma}\, d^2\zeta \, ,
   \ee
   where $\mu$ is the string tension and $\gamma$ the determinant of
   $\gamma_{\alpha\beta}=g_{\mu\nu}\pd_\alpha x^\mu \pd_\beta x^\nu$, 
   the pullback of the background metric (\ref{warped}) on the worldsheet.  
   The action (\ref{nambu}) enjoys 2D worldsheet diffeomorphism invariance, 
   which can be used to fix the gauge by imposing two conditions on 
   the worldsheet coordinates.  For our discussion it will be convenient 
   to work in the \emph{transverse temporal gauge}: 
   \be\label{gauge}
     \zeta^0=t, \quad \dot x^\mu x_\mu^{\prime}=0 \, ,
   \ee
   where a dot (resp. prime) denotes differentiation with respect to 
   the timelike (resp. spacelike) worldsheet coordinate $\zeta^0$ 
   (resp. $\zeta^1$).  This gauge choice imposes that $\dot x$ is 
   normal to the string, allowing an interpretation in terms of the
   physically relevant transverse string velocity, while identifying 
   worldsheet and background times.
    
   The equations of motion derived from the action (\ref{nambu}) in
   this gauge are: 
   \be\label{eom_expand} 
    \frac{\partial}{\partial t}\left(\frac{\dot x^{\mu}{x^{\prime}}^2} 
    {\sqrt{-\gamma}}\right) + \frac{\partial}{\partial \zeta} \left( 
    \frac{x^{\prime \mu}\dot x^2}{\sqrt{-\gamma}}\right) + \frac{1} 
    {\sqrt{-\gamma}} \Gamma^{\mu}_{\nu\sigma}\left({x^{\prime}}^2 
    \dot x^{\nu} \dot x^{\sigma} + \dot x^2  x^{\prime \nu} x^{\prime  
    \sigma}\right) = 0 ,  
   \ee 
   with $\mu,\nu,\sigma$ running from $0$ to $D$.  In the following 
   we shall use the notation: 
   \be\label{index_not} 
    \begin{array}{llrcl} 
     \mu,\nu=0,1,2,...,D& & & &                        \\ 
     i,j=1,2,3           &,&{\bf x}&\equiv& x^i       \\
     \ell,m=4,5,...,D    &,&{\bf l}&\equiv& x^\ell\,. \\ 
    \end{array}
   \ee     
   The Christoffel symbols of the metric (\ref{warped}) are: 
   \be\label{christ} 
    \begin{array}{lll} 
     \Gamma^0_{00}=\frac{\dot N}{N} & \Gamma^i_{00}=0 &  
       \Gamma^\ell_{00}=-\frac{1}{4}N^2\frac{h_{,\ell}}{b^2h^2}  \\ 
     \Gamma^0_{0i}=0 & \Gamma^i_{0j}=\frac{\dot a}{a}\delta^i_j & 
       \Gamma^\ell_{0i}=0 \\ 
     \Gamma^0_{0\ell}=-\frac{1}{4}\frac{h_{,\ell}}{h} & \Gamma^i_{0\ell}=0 & 
       \Gamma^\ell_{0m}=\frac{\dot b}{b}\delta^\ell_m  \\  
     \Gamma^0_{ij}=N^{-2} a \dot a \delta_{ij} & \Gamma^i_{jk}=0 & 
       \Gamma^\ell_{ij}=\frac{1}{4}\frac{a^2 h_{,\ell}}{b^2h^2}\delta_{ij} \\ 
     \Gamma^0_{\ell m}=hN^{-2}b\dot b\delta_{\ell m} & \Gamma^i_{\ell m}=0 & 
       \Gamma^\ell_{mn}=\frac{1}{4h}(\delta_{\ell m} h_{,n} +  
       \delta_{\ell n} h_{,m} - \delta_{mn} h_{,\ell}) \\
     \Gamma^0_{i\ell}=0 & \Gamma^i_{j\ell}=-\frac{1}{4}\frac{h_{,\ell}}{h} 
       \delta_{ij} & \Gamma^\ell_{im}=0 \,\,. \\
    \end{array}  
   \ee    
   We define a scalar $\epsilon$, the string energy per unit
   coordinate length (per unit tension), by:
   \be\label{eps_warped}   
    \epsilon=\frac{{-x^{\prime}}^2}{\sqrt{-\gamma}}={\left(\frac{h^{-1/2} 
    a^2{{\bf x}^{\prime}}^2+h^{1/2}b^2{{\bf l}^{\prime}}^2}{h^{-1/2}N^2-  
    h^{-1/2}a^2{\dot{\bf x}}^2-h^{1/2}b^2{\dot {\bf l}}^2}  
    \right)}^{1/2}    
   \ee
   and note that due to the gauge choice $\gamma_{01}\equiv \dot x^\mu 
   x_\mu^\prime=0$ we also have $\dot x^2/\sqrt{-\gamma}=\epsilon^{-1}$.  
   With this notation, the $0$, $i$ and $\ell$ components of the 
   equation of motion (\ref{eom_expand}) become: 
   \bq 
    &&\dot\epsilon=-\epsilon \left\{\frac{\dot N}{N}+\frac{a\dot a}{N^2}  
      \left[\dot{\bf x}^2-{\left(\frac{{\bf x}^{\prime}}{\epsilon}\right)}^2
      \right]+h\frac{b\dot b}{N^2}\left[\dot{\bf l}^2-{\left(\frac{{\bf l}^
      {\prime}}{\epsilon}\right)}^2\right]-\frac{{\bf l}\cdot\nabla h({\bf  
      l})}{4h}\right\} 
      \label{eom_eps_warp}\\
    &&\ddot{\bf x}+\left\{\frac{2\dot a}{a}-N^{-2}\left\{ N\dot N+a\dot
      a \left[\dot{\bf x}^2 - {\left(\frac{{\bf x}^{\prime}}{\epsilon}
      \right)}^2\right]+hb\dot b \left[\dot{\bf l}^2-{\left(\frac{{\bf l}^
      {\prime}}{\epsilon}\right)}^2 \right] \right\} \right\}\dot{\bf x} 
      \nonumber \\
    &&\;\;\;\;\;\;\;\;\;\;\;\;\;\;\;\;+\frac{1}{4h}\left({\bf l}^\prime 
      \cdot\nabla h({\bf l})\right)\epsilon^{-2}{\bf x}^\prime={\left( 
      \frac{{\bf x}^{\prime}}{\epsilon}\right)}^{\prime}\epsilon^{-1} 
      \label{eom_x_warp}\\
    &&\ddot{\bf l}+\left\{ \frac{2\dot b}{b}-N^{-2}\left\{ N\dot N+a\dot
      a \left[\dot{\bf x}^2-{\left(\frac{{\bf x}^{\prime}}{\epsilon}
      \right)}^2\right]+hb\dot b\left[\dot{\bf l}^2-{\left(\frac{{\bf l}^
      {\prime}}{\epsilon}\right)}^2\right]\right\}+3\frac{\dot{\bf l}\cdot 
      \nabla h({\bf l})}{4h}\right\}\dot{\bf l}\nonumber \\  
    &&\;\;\;\;\;\;\;\;\;\;\;\;\;\;\;\; -\frac{N^2\nabla h({\bf l})}{4b^2h^2} 
      +\frac{a^2\nabla h({\bf l})}{4b^2h^2}(\dot{\bf x}^2-\epsilon^{-2}{{\bf   
      x}^\prime}^2)-\frac{\nabla h({\bf l})}{4h}(\dot{\bf l}^2-\epsilon^{-2} 
      {{\bf l}^\prime}^2) \nonumber \\  
    &&\;\;\;\;\;\;\;\;\;\;\;\;\;\;\;\; -\frac{1}{2h}\left({\bf l}\cdot\nabla  
      h({\bf l})\right)\epsilon^{-2}{\bf l}^\prime={\left(\frac{{\bf l}^ 
      {\prime}}{\epsilon}\right)}^{\prime}\epsilon^{-1} \,. 
      \label{eom_l_warp}
   \eq    

   Comparing to the corresponding equations \cite{EDVOS} in the
   unwarped case $h({\bf l})=1$, one observes that the effect of warping 
   is to introduce factors of $h$ (in the $\dot b$ terms) and new 
   potential terms proportional to $\nabla h({\bf l})$.  This is in 
   agreement with the intuitive expectation that, since energy is minimised 
   at highly warped regions, there should be forces driving the string 
   towards those regions.  However, the dynamics by which this 
   localisation mechanism may operate has not been studied in detail.  
   The primary purpose of this paper is to explore the effect these 
   potential terms may have on string evolution.  In the next section 
   we will study this problem by considering the relevant macroscopic 
   velocity equations for simple warping potentials.

 \section{\label{VOS}Effect of Warping on String Evolution: A Simple Model} 

  In this section we shall study the effect of warping on the
  evolution of string networks, using a macroscopic velocity-dependent
  network model analogous to the models of Refs.~\cite{vos,vosk,EDVOS}.  
  The picture we have in mind is a network of strings produced at the 
  end of brane inflation \cite{DvalTye,DvalShafSolg,BMNQRZ,Garc-Bell, 
  JoStoTye1,KKLMMT}.  Assuming that reheating is efficient, the 
  strings can be taken to evolve in a radiation-dominated universe, 
  but there is also a compact manifold of internal dimensions, which 
  can have significant impact on the evolution of the network 
  \cite{JoStoTye2,EDVOS,PolchProb}.  In explicit constructions 
  \cite{KKLMMT}, the metric is a warped product, and the warping 
  factor is a function of the internal coordinates, giving rise 
  to `throats' of local minima in the internal manifold.  In this  
  section we will consider a simplified situation in which the  
  extra dimensions are toroidally compactified, so the metric is 
  of the form (\ref{warped}).  This will allow us to write down  
  explicit evolution equations for the strings and obtain numerical  
  solutions given a choice of warping factor.    
    
  The idea is to average the equations of motion 
  (\ref{eom_eps_warp})-(\ref{eom_l_warp}) over a network of Nambu-Goto 
  strings to obtain macroscopic evolution equations for the 
  root-mean-squared (rms) velocities of string segments.  Let us first 
  consider the $\ddot {\bf x}$ equation (\ref{eom_x_warp}).  This 
  differs from the corresponding unwarped spacetime equation 
  \cite{EDVOS} in two ways: first, there is a factor of $h$ multiplying 
  $b\dot b(\dot{\bf l}^2-{{\bf l}^{\prime}}^2 \epsilon^{-2})$, and 
  second, there is the new term: 
  \be\nonumber 
   \frac{1}{4h}\left({\bf l}^\prime\cdot\nabla h({\bf l})\right) 
   \epsilon^{-2}{\bf x}^\prime\,. 
  \ee 
  For stabilised extra dimensions we have $\dot b=0$, so the first of 
  the above terms is zero in both unwarped and warped backgrounds.  
  Further, in order to obtain the macroscopic velocity equation, one 
  has to dot with $\dot{\bf x}$ and average over the string network, 
  so the second term yields a contribution proportional to $\langle 
  {\bf x}^\prime\cdot\dot{\bf x}\rangle$, where the angled brackets 
  denote `energy weighted averaging' obtained by integrating over 
  the worldsheet with weighting function $\epsilon$, and normalising 
  with respect to the total string energy.  But, since the 3-vectors 
  $\dot{\bf x}$ and ${\bf x}^\prime$ are uncorrelated, one expects 
  $\langle{\bf x}^\prime\cdot\dot{\bf x}\rangle$ to randomly change 
  sign with no large-scale correlations as one moves along the string, 
  and so this term gives zero when averaged over large scales. 
  This has been tested numerically in the case of a 3+1 FLRW model 
  in Ref.~\cite{EDVOS}.  Finally, there is an implicit dependence on 
  the warp factor through the modified definition of $\epsilon$ in 
  equation (\ref{eps_warped}), which, however, disappears when one 
  moves to physical variables.  Indeed, defining the rms (peculiar) 
  string velocities: 
   \be\label{v_x_warp}
    v_x^2=\left\langle\left(\frac{h^{-1/4}a {\rm d}{\bf x}}{h^{-1/4}N
    {\rm d}t} \right)^2 \right\rangle = \left\langle\left(\frac{{\rm d}
    {\bf x}}{{\rm d} \tau}\right)^2 \right\rangle \equiv \langle\dot  
    {\bf x}^2 \rangle
   \ee
  and
   \be\label{v_l_warp}
    v_{\ell}^2=\left\langle\left(\frac{h^{1/4}{\rm b d}{\bf l}}{h^{-1/4}
    N{\rm d}t}\right)^2\right \rangle = \left\langle\left(\frac{h^{1/2}
    b {\rm d} {\bf l}}{a {\rm d}\tau} \right)^2\right\rangle \equiv 
    \langle h b^2 {\dot {\bf l} }^2 / a^2 \rangle \, , 
   \ee
  where `conformal' time $\tau$\footnote{In this section we use the 
  notation $\dot{}\equiv\frac{\rm d}{{\rm d}\tau}$.} corresponds to 
  the slicing $N\!=\!a$, the evolution equation for the 3-dimensional 
  velocity $v_x$ in terms of the `proper' time ${\rm d}s=h^{-1/4}N 
  {\rm d}t$ is identical to 
  that of the unwarped case, namely: 
  \be\label{v_xdt}
    v_x \frac{{\rm d} v_x}{{\rm d}s}=\frac{k_x v_x}{R}(1-v^2)-\left(
    2-w_\ell^2\right) \frac{1}{a}\frac{{\rm d}a}{{\rm d}s}(1-v^2) 
    {v_x}^2 - \frac{1}{a} \frac{{\rm d}a}{{\rm d}s} {v_\ell}^2{v_x}^2 
    \, .  
   \ee
  Here, $v^2=v_x^2+v_\ell^2$ and $w_\ell$ is a string orientation 
  parameter
   \be\label{wl}
    w_\ell=\left\langle \frac{ h b^2 {{\bf l}^{\prime}}^2 }  
    {a^2 {{\bf x}^{\prime}}^2 + h b^2 {{\bf l}^{\prime}}^2 }
    \right \rangle^{1/2} , 
   \ee
   quantifying the degree to which the strings lie in the extra
   dimensions ${\bf l}$.  The 3-momentum parameter $k_x$ is defined 
   by
   \be\label{k_x}
    \frac{k_x v_x (1-v^2)}{R}=\left\langle \dot {\bf x}\cdot  {\bf u}
    \left( 1-\dot{\bf x}^2-\frac{h b^2 \dot{\bf l}^2 }{a^2}\right) 
    \right\rangle\,,
   \ee
  where ${\bf u}$ is the physical curvature vector and $R$ the average
  radius of curvature of the string network (see Ref.~\cite{EDVOS} for 
  details). 

  We now consider the $\ddot{\bf l}$ equation (\ref{eom_l_warp}).  This 
  contains several extra terms, some of which survive after averaging 
  over the worldsheet.  In particular there are terms proportional to 
  $\nabla h({\bf l})$, which can be thought of as a force driving the 
  strings towards the minima of the warping potential.  Indeed, for 
  a static string configuration, the worldsheet action reduces to 
  a potential\footnote{There is also a dependence on the dilaton,  
  which is ignored here (see later discussion).} $V({\bf l})=\mu  
  h^{-1/2}$ \cite{PolchProb} and the corresponding force $F=-\nabla  
  V({\bf l})$ is proportional to $\nabla h({\bf l})h^{-3/2}$.  An  
  equation like (\ref{v_xdt}) only yields information about the time  
  evolution of the magnitude of the velocity but, here, we are also 
  interested in its direction.  We will thus seek to construct a 
  vector equation for the internal velocities ${\bf v}_{\ell}$.   
       
  For simplicity, we will choose a special configuration in which 
  the strings are oriented normally to the extra dimensions, that is 
  we will set ${\bf l}^\prime=0$.  In this way we eliminate effects 
  arising from string curvature in the extra dimensions (the right 
  hand side of (\ref{eom_l_warp})) as well as corrections proportional 
  to $w_{\ell}$ (see equation (\ref{v_xdt})), concentrating only  
  on the effects of the warped background.  One may worry that this
  choice could suppress effects that might be relevant in the following 
  analysis, but it turns out that this is not the case.  The effects 
  of string bending in the extra dimensions have been studied in
  Refs.~\cite{EDVOS,intprob} and, on macroscopic\footnote{In this 
  context `macroscopic' refers to scales greater than the string 
  correlation length.} scales, can be described by a non-zero 
  $w_{\ell}$ and an effective renormalisation of the string tension 
  $\mu_{\rm eff}>\mu$, both of which will not be important in the 
  following.  On the other hand, on small scales that are relevant 
  in the present study, this bending can produce string velocities 
  in the extra dimensions.  However, these will simply add to string 
  kinetic energies and can only strengthen our conclusions, which 
  will be based on the absence of an efficient damping mechanism in 
  these dimensions.  

  By concentrating on this special configuration ${\bf l}^\prime=0$, 
  equation (\ref{eom_l_warp}) simplifies considerably and this will 
  allow us to study the dynamics of strings near a minimum of the  
  warping potential \cite{thesis}.  We define the physical velocity 
  in the extra dimensions as the $(D-3)$-vector
  \be\label{v_l_warp_vect} 
   {\bf v}_{\ell}=h^{1/2} b \dot{\bf l} / a \,,  
  \ee
  which, due to the chosen string orientation, does not depend on 
  the spacelike worldsheet coordinate $\zeta$.  Then, by forming  
  $\dot{\bf v}_{\ell}$ and using the equations of motion  
  (\ref{eom_eps_warp}), (\ref{eom_l_warp}) we obtain: 
  \be\label{v_l_warp_vecteqn} 
   \frac{{\rm d}{\bf v}_{\ell}}{{\rm d}s}=-\left[\frac{1}{a}\frac{{\rm d}
   a}{{\rm d}s}\left(1-2v_x^2-v_\ell^2\right)+\frac{{\bf v}_{\ell}\cdot
   \nabla h({\bf l})}{4bh^{5/4}}\right]{\bf v}_{\ell}+\left(2-2 v_x^2 
   \right) \frac{\nabla h({\bf l})}{4bh^{5/4}} \, . 
  \ee
 
\begin{figure} 
   \includegraphics[height=2.5in,width=2.8in]{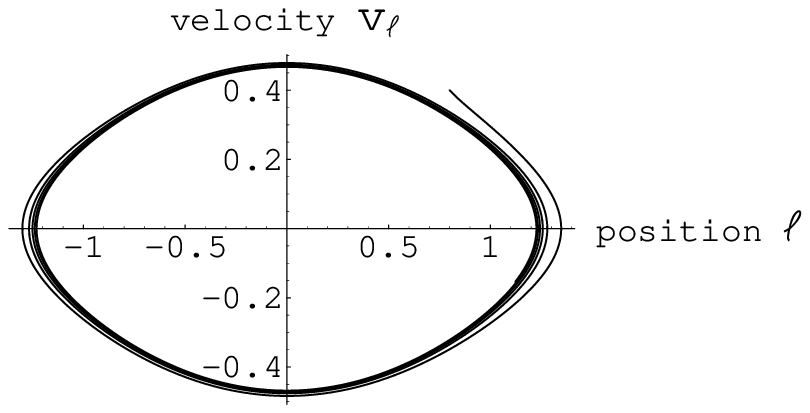}
   \includegraphics[height=2.5in,width=2.8in]{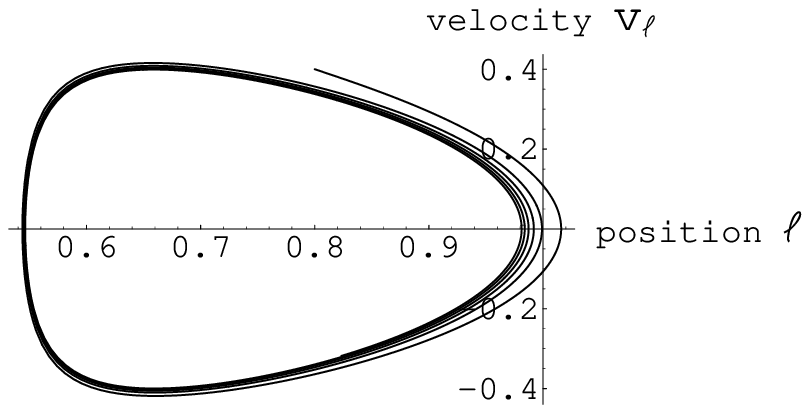}
   \caption{\label{vlvec} String trajectory in two-dimensional 
            phase space $(\ell,v_{\ell})$, i.e. in the case of a single 
            extra dimension $\ell$, assuming a constant 3D velocity 
            $v_x$.  The two plots correspond to different warping
            potentials, with warp factors $h(\ell)=A-B\tanh^2(\ell)$ 
            (left) and $h(\ell)=[A+B\ln(|\ell|)]/\ell^4$ (right).  
            Starting at a distance away from the potential minimum 
            (located at $\ell=0$ in the former case and $\ell=0.8$ 
            in the latter) 
            and with initial velocity towards it, the string 
            oscillates around the tip, but the motion is only 
            weakly damped by Hubble expansion.}
  \end{figure}

\begin{figure}
   \includegraphics[height=2.5in,width=2.8in]{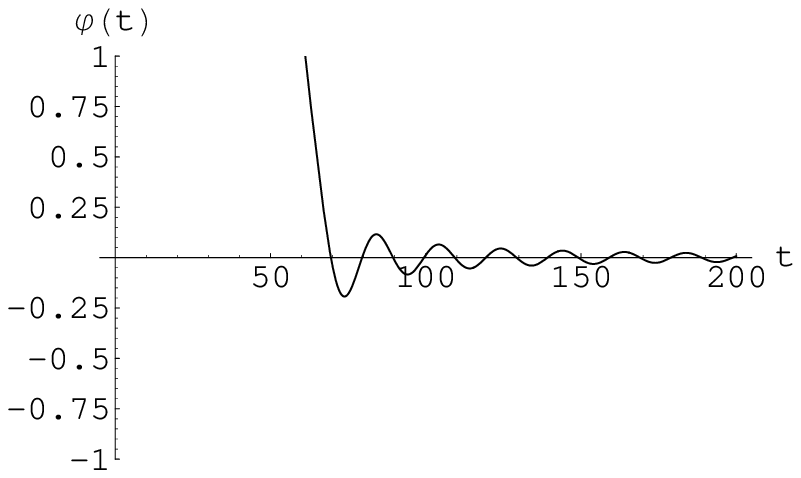}
   \includegraphics[height=2.5in,width=2.8in]{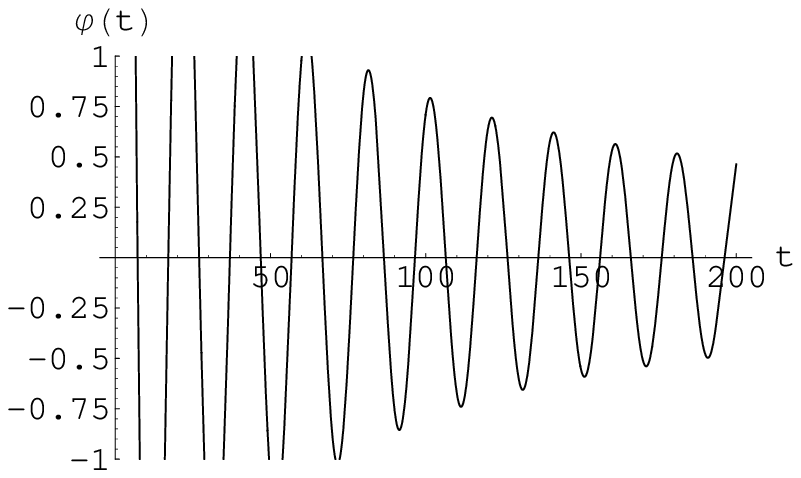}
   \caption{ 
            \label{damping} Effect of Hubble damping on scalar 
            field oscillations $\varphi(t)$ in both scalar field 
            dominated and radiation dominated cosmology.  In the 
            scalar field dominated case (left), Hubble damping 
            has approximately constant magnitude, but in radiation 
            domination  
            (right) cosmological friction is much less efficient 
            as it scales like $t^{-1}$.  In the case of strings  
            the situation is more dramatic, because the Hubble 
            term also comes with a factor of $1-2v_x^2-v_\ell^2$  
            (see equation (\protect\ref{v_l_warp_vecteqn})) and 
            so, as 3D velocities evolve towards their scaling 
            value, this term becomes zero.}
  \end{figure}

  We can use this equation to study the dynamics of a straight string
  moving in a warping potential $h({\bf l})$.  Fig.~\ref{vlvec} shows 
  the relevant phase diagram for two simple choices of warping potentials, 
  namely $h({\bf l})=A-B\tanh^2({\bf l})$ and $h({\bf l})=[A+B\ln(|{\bf 
  l}|)]/{\bf l}^4$, in the simplest case of one extra dimension and 
  assuming a constant 3-dimensional velocity $v_x$.  As expected, the 
  strings are driven towards the minimum of the potential at the centre, 
  but the damping provided by Hubble expansion is too weak to guarantee  
  that they actually reach it.  Instead, equation (\ref{v_l_warp_vecteqn}) 
  suggests a picture in which the string oscillates around the minimum, 
  rather than quickly falling on it and stabilising.  In view of the 
  results of Ref.~\cite{EDVOS} this is not surprising: there, it was  
  found that Hubble damping couples very weakly to the extra dimensional 
  velocities and is generally insufficient to cause significant 
  redshifting of velocities in the internal dimensions.  This may seem 
  to contradict the intuition one has from inflation, where oscillations 
  of the inflaton around the minimum of the potential at the end of  
  inflation are efficiently damped.  Here the situation is similar, as  
  the internal position of the string corresponds to a scalar field in 
  four dimensions.  However, unlike the case of inflation where cosmology  
  is scalar field dominated and the damping is efficient, here the  
  oscillations take place during radiation domination, so the damping 
  term is much weaker and decays as $t^{-1}$ (see Fig.~\ref{damping}).      

\begin{figure}
   \includegraphics[height=2.3in,width=2.6in]{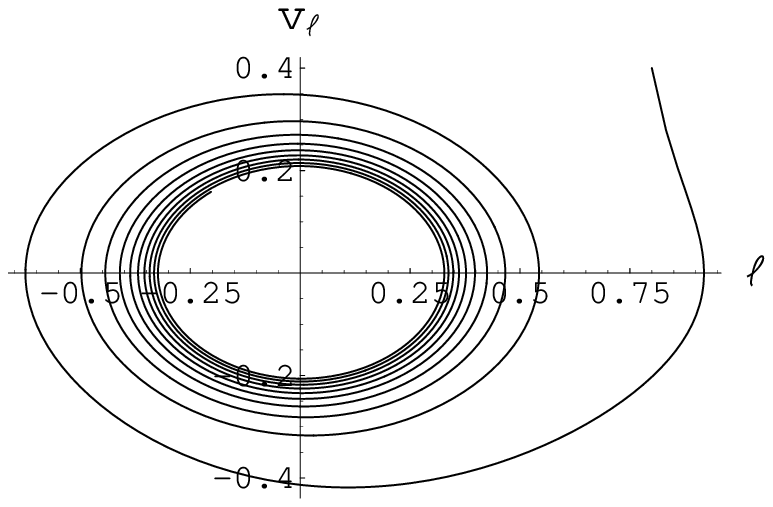}
   \includegraphics[height=2.3in,width=2.6in]{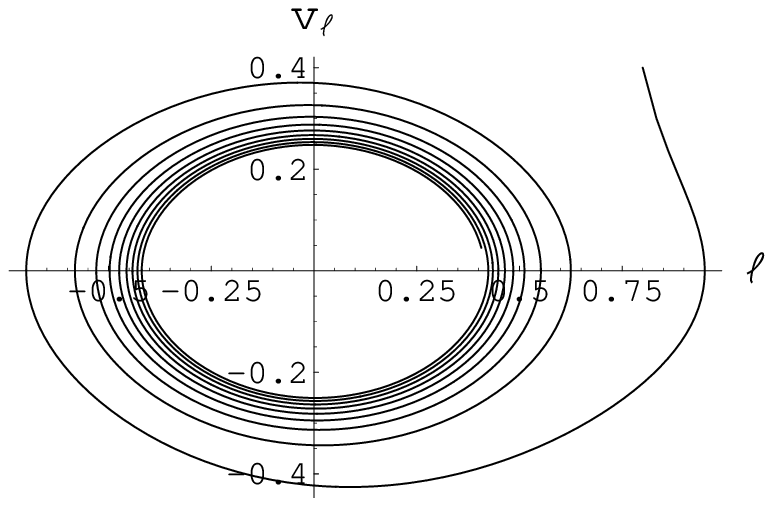}
   \includegraphics[height=2.3in,width=2.6in]{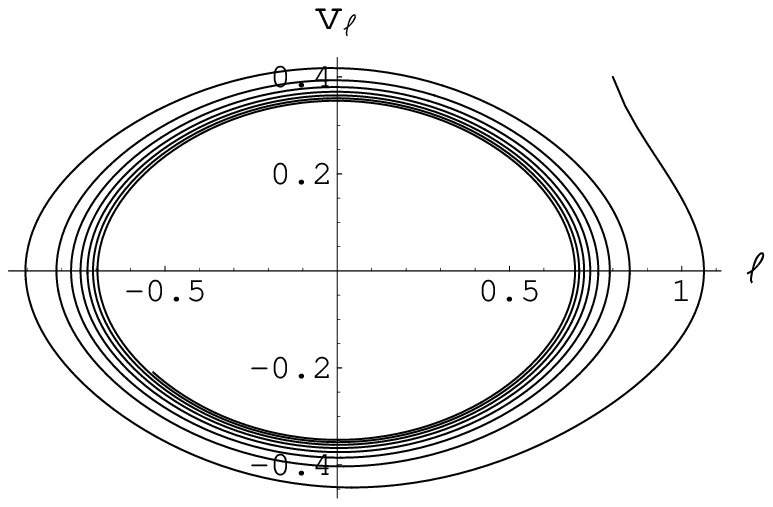}
   \includegraphics[height=2.3in,width=2.6in]{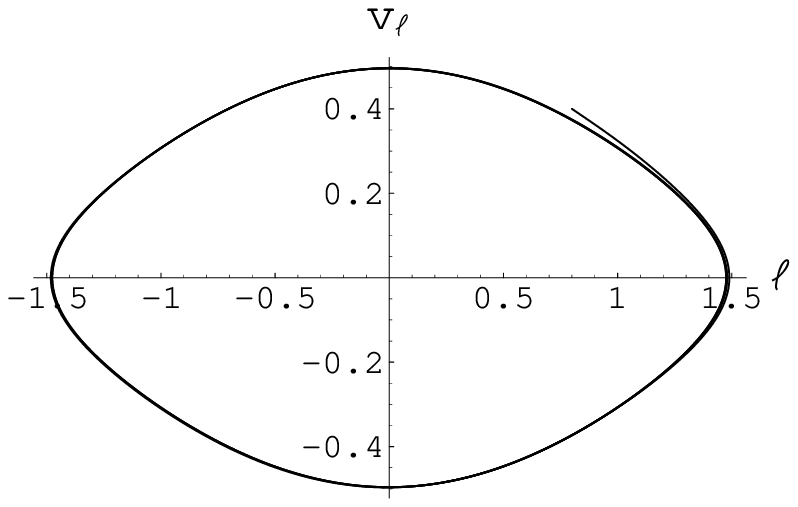}
   \caption{\label{vx_dependence}  Effect of 3D velocity $v_x$  
            on the orbits of the string in $(\ell,v_{\ell})$ phase 
            space, for the case of one extra dimension $\ell$.  
            The warping factor is taken to be of the form 
            $h(\ell)=A-B\tanh^2(\ell)$.  Treating the 3D velocity 
            as a constant parameter, each plot corresponds 
            to a differnt value of $v_x$.  From top left to 
            bottom right: $v_x=0$, $0.3$, $0.5$ and $0.64$.  
            Clearly, as one increases $v_x$, Hubble damping 
            becomes less and less important, since the 
            coefficient of the friction term in equation 
            (\protect\ref{v_l_warp_scaleqn}) decreases.  As 
            a result, the orbit is more stable for larger 
            values of $v_x$.  To model this effect, the 3D 
            velocity should be treated as a dynamical variable 
            rather than a constant parameter.}
  \end{figure}

\begin{figure}
   \includegraphics[height=2.3in,width=2.6in]{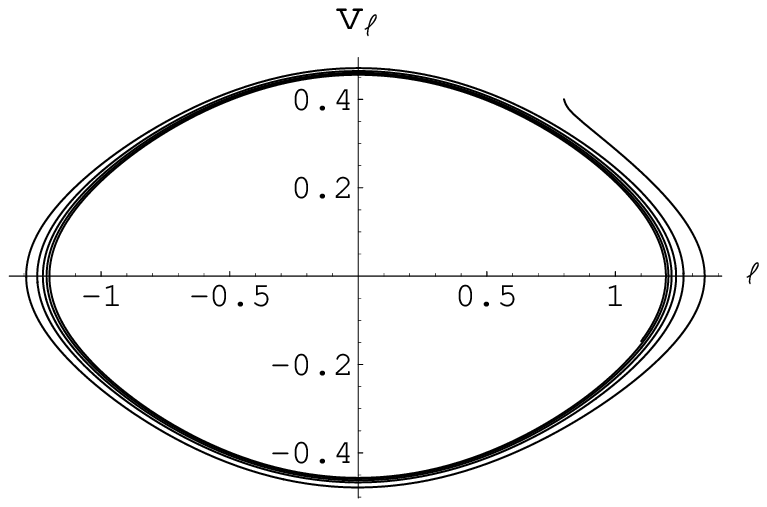}
   \includegraphics[height=2.3in,width=2.6in]{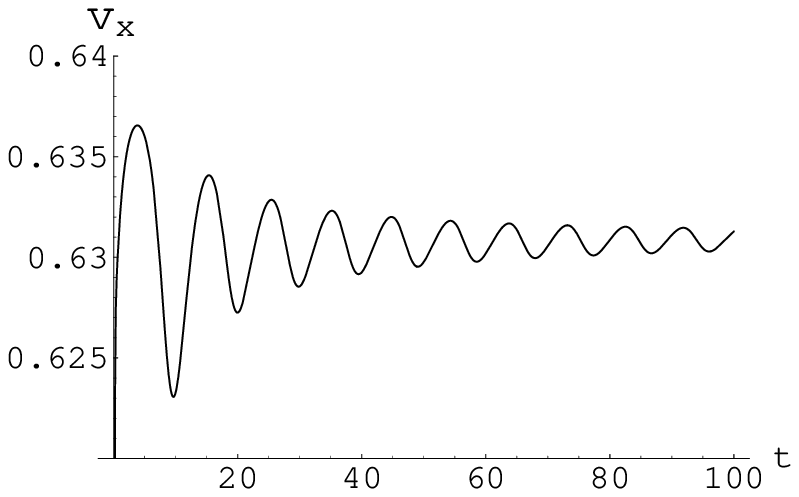}
   \includegraphics[height=2.3in,width=2.6in]{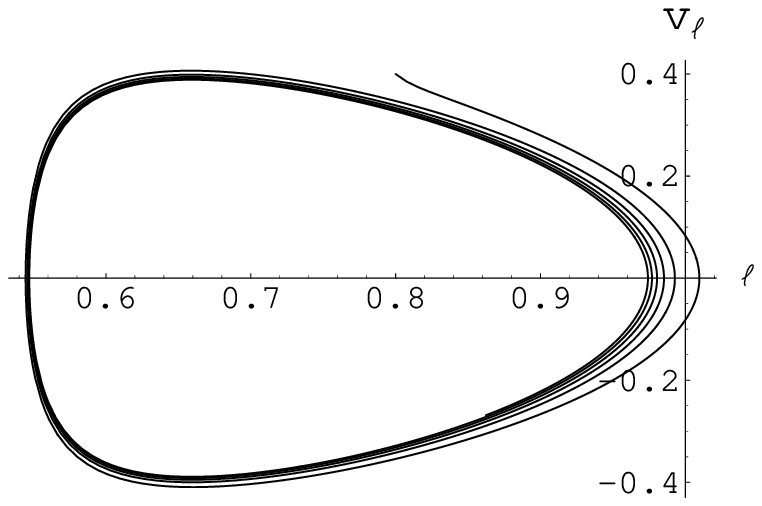}
   \includegraphics[height=2.3in,width=2.6in]{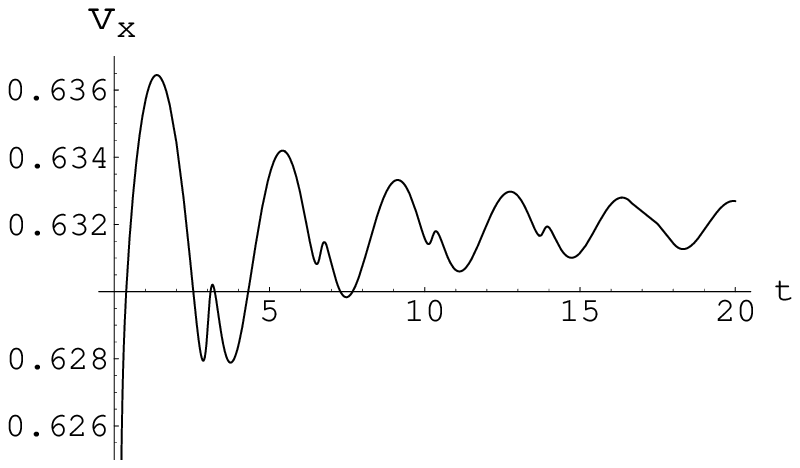}
   \caption{\label{vlvecvx}  String trajectories (left)  
            near the warped minimum in two-dimensional phase 
            space $(\ell,v_{\ell})$ for the same warping potentials 
            and initial conditions as in Fig.~\protect\ref{vlvec}, 
            but now treating $v_x$ as a dynamical variable.  The 
            3D velocity $v_x$ oscillates due to its coupling 
            to $v_\ell$ (see equation (\protect\ref{v_xdt_w_l_zero})), 
            but this effect is small (right).  The upper plots 
            are for warping factor $h(\ell)=A-B\tanh^2(\ell)$, while 
            the lower ones for $h(\ell)=[A+B\ln(|\ell|)]/\ell^4$.}
  \end{figure}

  Let us briefly discuss the effect of the 3-dimensional velocity 
  $v_x$ on the orbits of the string in $({\bf l}, {\bf v}_\ell)$ 
  phase-space.  In the case of a single compact dimension $\ell$, 
  equation (\ref{v_l_warp_vecteqn}) reads:    
  \be\label{v_l_warp_scaleqn}
   \frac{{\rm d} v_{\ell}}{{\rm d}s}=-\frac{1}{a}\frac{{\rm d}
   a}{{\rm d}s}\left(1-2v_x^2-v_\ell^2\right) v_\ell+\left(2-2 v_x^2-
   v_\ell^2 \right)\frac{\nabla h(\ell)}{4bh^{5/4}} \, , 
  \ee
  where $v_\ell$ can also take negative values.  Note that the first 
  term, corresponding to Hubble damping, comes with a coefficient of 
  $(1-2v_x^2-v_\ell^2)$ which can be much smaller than 1 for $v_x^2 
  \lesssim 1/2$.  The strength of this term depends on the magnitude 
  of $v_x$ and decays as $t^{-1}$ as the universe expands.  On the 
  other hand, the potential term comes with a coefficient of $(2 
  -2v_x^2-v_\ell^2)$, which is always greater than unity due to 
  the constraint $v^2\le 1/2$.  This term is not diluted by cosmic 
  expansion.  The evolution is therefore dominated by this potential 
  term but there is also a transient Hubble damping effect, which 
  operates for a few Hubble times until it effectively dies away, 
  and whose strength depends on the 3-dimensional velocity $v_x$ 
  (Fig.~\ref{vx_dependence}).  It is therefore important to treat 
  $v_x$ as a dynamical variable rather than a constant parameter.  
  Also, as we saw, the evolution of $v_x$ is governed by equation 
  (\ref{v_xdt}), which depends on $v_\ell$, and thus the assumption 
  of constant $v_x$ is at best an approximation.  We will therefore 
  couple equation (\ref{v_l_warp_scaleqn}) to the evolution equation 
  for the rms $v_x$, which, for the special orientation we chose  
  ($w_\ell=0$), becomes:  
  \be\label{v_xdt_w_l_zero}
    \frac{{\rm d} v_x}{{\rm d}s}=\frac{k_x}{R}(1-v^2)-
    \frac{1}{a}\frac{{\rm d}a}{{\rm d}s}(2-2v_x^2-v_\ell^2) v_x
    \, .
  \ee
  This yields very interesting dynamics, with $v_x$ exhibiting damped 
  oscillations around its average value (Fig.~\ref{vlvecvx}).  This 
  effect, however, is too small to be of observational interest and is 
  further suppressed by cosmic expansion.  Initially, the string oscillates 
  around the warped minimum, being (weakly) damped by Hubble expansion, 
  while its 3-dimensional velocity $v_x$ is modulated by the oscillation. 
  After a few revolutions, the damping dies away and the phase-space 
  orbit stabilises.    

\begin{figure}
   \includegraphics[height=1.5in,width=1.6in]{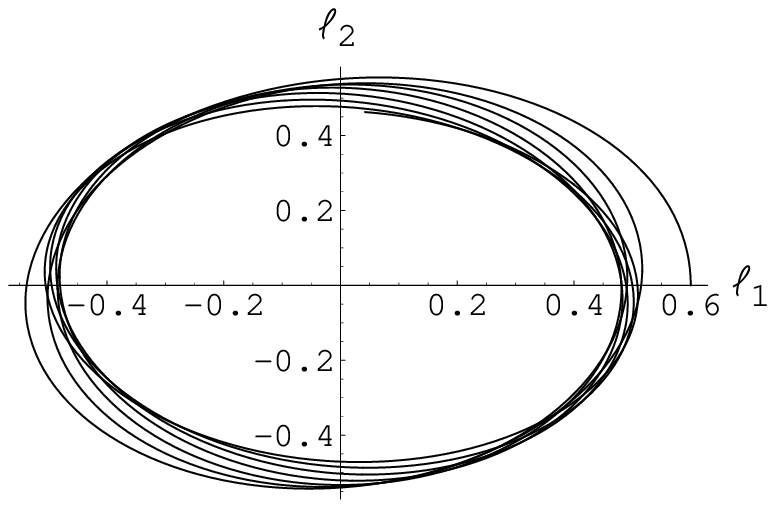}
   \includegraphics[height=1.5in,width=1.6in]{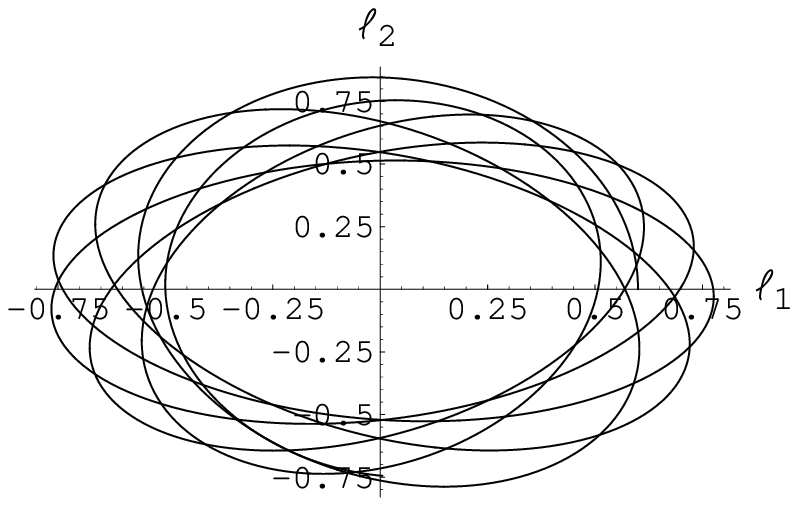}
   \includegraphics[height=1.5in,width=1.6in]{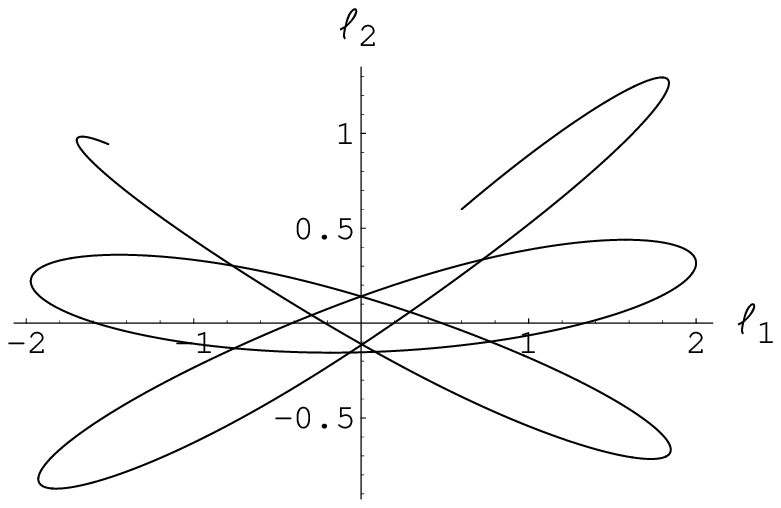}
   \includegraphics[height=1.5in,width=1.6in]{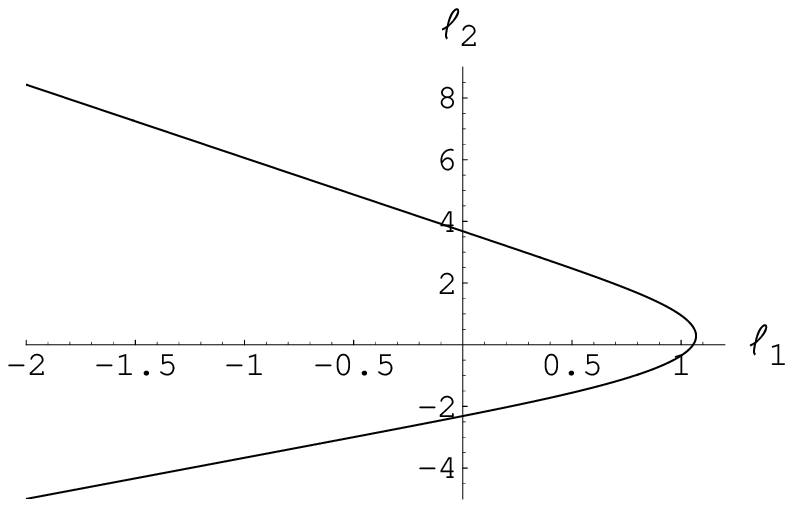}
   \includegraphics[height=1.5in,width=1.6in]{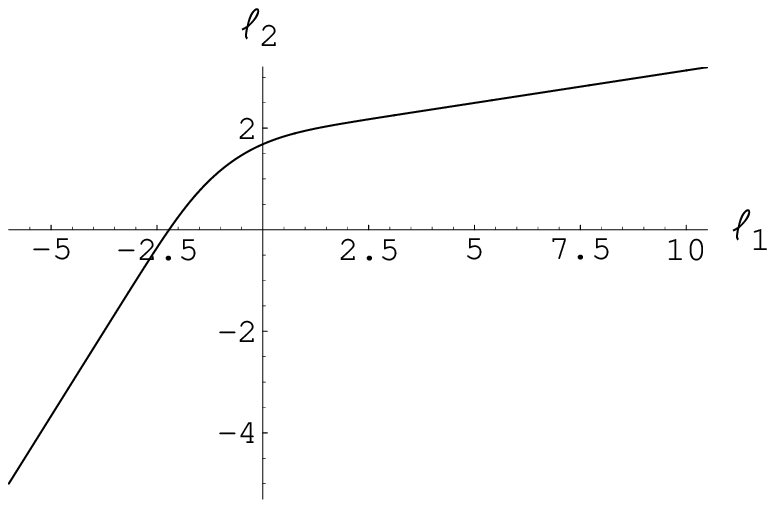}
   \includegraphics[height=1.5in,width=1.6in]{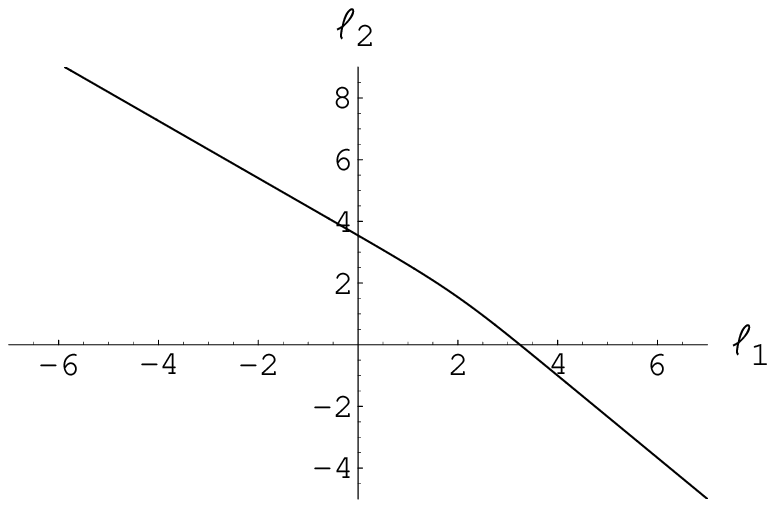}
   \caption{\label{2_extra_dims} A sample of string trajectories in
            physical space, in the case of two extra dimensions 
            $\ell_1$ and $\ell_2$.  Different trajectories correspond 
            to different initial conditions for the position 
            ${\bf l}=(\ell_1,\ell_2)$ 
            and velocity ${\bf v}_{\ell}=(v_{\ell 1},v_{\ell 2})$. 
            In all cases the warping factor is $h({\bf l})
            =A-B\tanh^2(|{\bf l}|)$.  The possibilities (depending 
            on the initial conditions) include deflections, bounces 
            and bound orbits with negligible Hubble friction.} 
  \end{figure}

  In this simplest case of a single extra extra dimension that we have 
  studied so far, the string has to pass through the minimum in coordinate 
  space in every cycle, but in the presence of more internal dimensions   
  there will generally be a non-zero impact parameter.  One naturally 
  expects that angular momentum conservation will lead to 
  deflecting/bouncing orbits around the tip of the warped throat.  
  This will be studied in some detail in the next section.  For now, 
  we simply plot a sample of string trajectories (now in physical 
  space), obtained by solving equations (\ref{v_l_warp_vecteqn}) 
  and (\ref{v_xdt_w_l_zero}) in the 
  case of two internal dimensions (Fig.~\ref{2_extra_dims}).  As 
  shown, depending on the initial conditions, the string is deflected 
  or bounces back, and can either escape from the throat or enter a 
  series of bounces around the tip.           

  The key point of this section, which is the central idea of the 
  present paper, is that the classical evolution does not possess a 
  strong damping term to guarantee that the strings quickly migrate 
  to the tip of the warping throat and stabilise there.  Instead,  
  depending on the velocity and impact parameter, a string passing 
  near a warped throat will generally experience a mere deflection or 
  a bounce around the potential minimum.  Note that, in the above, we 
  have ignored any possible dependencies on the dilaton $\Phi({\bf l})$.  
  In the case of $(p,q)$-strings \cite{PolchStab} for example, taking 
  into account the dilaton dependence gives rise to a factor of 
  $(p^2+q^2 e^{-2\Phi({\bf l})})^{1/2}$ \cite{PolchProb} in the 
  potential $V({\bf l})$.  This could lead to a dependence of the 
  minimum on $p$ and $q$, hence on the type of string, with potentially 
  important implications for the intercommuting properties of 
  different types of strings.  However, for the strongly warped 
  backgrounds one is usually interested in (e.g. \cite{KKLMMT}),  
  the variation of the dilaton is negligible and thus we have 
  safely ignored this effect in our discussion.

 \section{\label{IIB}IIB Compactifications} 

 In the previous section, we considered string motion in a 
 background that was a warped product of an expanding FLRW 
 universe with a compact internal manifold.  Here, we would 
 like to make contact with explicit constructions in string 
 theory, in particular type IIB fluxed compactifications 
 that have been used to realise cosmological inflation 
 \cite{KKLMMT,Pajer,BDKMcA}.  These typically involve a metric  
 that is a warped product of Minkowski spacetime with a Calabi-Yau  
 manifold, and time-dependence of the background arises in 
 the effective 4D description.  Indeed, in the effective 
 theory, a number of scalar fields appear, which can couple 
 to the 4D metric.  In the constructions of \cite{KKLMMT,BDKMcA}, 
 all scalar fields are dynamically stabilised apart from one, 
 corresponding to the position of a mobile $D$-brane in the 
 10D picture, which plays the role of the inflaton in the 
 effective theory.       

 The Einstein frame metric takes the following general form:
  \be\label{metric_E}
   ds^2 = h^{-1/2}(l)\eta_{\mu\nu} {\rm d}x^\mu {\rm d}x^\nu 
   -h^{1/2}(l) g_{\ell m} {\rm d}l^\ell {\rm d}l^m ,  
  \ee 
 where $\eta_{\mu\nu}$ is the 4D Minkowski metric and $g_{\ell m}$ 
 the metric on the internal Calabi-Yau space.  The warp factor 
 $h$ depends only on the internal coordinates $l$.  Explicit 
 solutions of this type exist, for example Klebanov-Tseytlin 
 (KT)~\cite{KlebTseyt} and Klebanov-Strassler (KS)~\cite{KlebStras}  
 geometries, but, for the general discussion that follows, we 
 leave the exact geometry unspecified.  We will assume, however, 
 as in the above solutions, that the internal manifold has a 
 group of angular symmetries allowing us to define a radial 
 coordinate $r\equiv l^r$, and that the warp factor $h$ depends 
 only on this radial coordinate.   
   
 A moving string on this warped background is described by the 
 Nambu-Goto action (\ref{nambu}) with metric (\ref{metric_E}).
 The motion of $D$-branes in warped backgrounds has been 
 well-studied~\cite{Kutasov,KacMcAl,Germani,EaGrTaZa}.  The 
 solutions found in those cases include deflections, bounces, 
 and bound  orbits, like in the previous section.  Unlike the 
 case of $D3$-branes, which are spacetime-filling and only have 
 velocities in the internal space, strings are rather different, 
 as there are also two transverse directions where the string can 
 move, giving rise to 3D velocities that can dynamically interfere 
 with the internal ones (see previous section).  We consider this 
 case in more detail below.     

 Defining the radial direction $r\equiv l^r$, we write the internal 
 metric as: 
 \be\label{int_metric}
  g_{\ell m} {\rm d}l^\ell {\rm d}l^m = g_{rr} {\rm d}r^2 
  + g_{\theta\phi} {\rm d}l^\theta {\rm d}l^\phi ,
 \ee
 where the indices $\theta$ and $\phi$ run over the angular internal 
 coordinates.  The internal speed of the string is then $\dot l^2 
 = g_{rr} \dot r^2 + g_{\theta\phi} \dot l^\theta \dot l^\phi$, and 
 the action, in the transverse temporal gauge, reads: 
 \be\label{action_sym} 
  S=-\mu \int h^{-1/2}(r) \sqrt{\left[1-\dot{\bf x}^2 - h(r) 
  (g_{rr} \dot r^2 + g_{\theta\phi} \dot l^\theta \dot 
  l^\phi)\right]\left({\bf x}^{\prime 2}+h(r)g_{\ell m} l^{\prime\ell} 
  l^{\prime m}\right)} \,\, d^2\zeta \, .  
 \ee     
 As in the previous section, we will consider the special 
 configuration in which the string has $l^\prime=0$.  As we
 have chosen the angular coordinates $l^\phi$ to correspond
 to spacelike Killing vectors, the following momenta are 
 conserved: 
 \be\label{momenta} 
  \pi_\phi \equiv \frac{\partial{\cal L}}{\partial\dot l^\phi} = 
  \frac{\mu\sqrt{{\bf x}^{\prime 2}}}{\sqrt{1-\dot{\bf x}^2 - 
  h(r) (g_{rr} \dot r^2 + g_{\theta\omega} \dot l^\theta \dot   
  l^\omega)}} h^{1/2}(r) g_{\phi\theta} \dot l^\theta \, .
 \ee 
 Also, time translational invariance implies that the energy 
 \be\label{energy} 
  {\cal E}\equiv {\bf p}\cdot\dot{\bf x} + \rho \dot r + \pi_\phi  
  \dot l^\phi - {\cal L} = \frac{\mu h^{-1/2}(r)\sqrt{{\bf x}^{\prime 
  2}}}{\sqrt{1-\dot{\bf x}^2 - h(r) (g_{rr} \dot r^2 + g_{\theta\phi} 
  \dot l^\theta \dot l^\phi)}}
 \ee
 is conserved, where ${\bf p}$ and $\rho$ are the canonical momenta 
 associated to ${\bf x}$ and $r$ respectively.  Then, defining 
 \be\label{Pi_of_r} 
  \Pi^2(r) \equiv g^{\theta\phi} \pi_\theta \pi_\phi \, ,
 \ee  
 we can write 
 \be\label{E_of_Pi} 
  {\cal E}=\mu\sqrt{{\bf x}^{\prime 2}} h^{-1/2}(r)\left(\frac{1+ 
  \Pi^2(r)/\mu^2 {\bf x}^{\prime 2}}{1-\dot{\bf x}^2-h(r)g_{rr} 
  \dot r^2}\right)^{1/2} \, , 
 \ee
 or equivalently: 
 \be\label{rdot2} 
  \dot r^2 = \frac{g^{rr}}{h(r)} \left[ 1 - \left( \frac{\mu^2
  {\bf x}^{\prime 2}+\Pi^2(r)}{h(r){\cal E}^2} + \dot{\bf x}^2 \right) 
  \right] \, .
 \ee 
 
 For the setup we are interested in, the strings are macroscopic 
 in the Minkowskian directions and homogeneous over the short 
 length-scales relevant to the warping scale in the internal 
 dimensions.  Further, as we saw in the previous section, the 
 coupling between 3D and internal velocities is weak, 
 so one could take constant $\dot{\bf x}^2$ as an approximation.  
 Here, we will consider the special case of a straight string 
 ${\bf x}^{\prime 2}=1$.  Translational invariance along the 
 transverse string directions implies that the corresponding 
 momenta $\bf p$ are conserved, and we can write: 
 \be\label{rdot2_simple}
  \dot r^2 = \frac{g^{rr}}{h(r)} \left( 1 - \frac{\mu^2
  +\Pi^2(r)+h(r){\bf p}^2}{h(r){\cal E}^2} \right) \, .
 \ee
 
 The right-hand-side is a function of the radial coordinate only, 
 and so equation (\ref{rdot2_simple}) describes the one-dimensional 
 motion of a particle in an effective potential 
 \be\label{eff_pot} 
  \dot r^2 + V_{\rm eff}(r) = 0 \, , 
 \ee
 where $V_{\rm eff}(r)$ is minus the right-hand-side of 
 (\ref{rdot2_simple}).  Physical motion is restricted to regions  
 where $V_{\rm eff}\le 0$, and the zeros of the effective potential 
 correspond to turning points, where the string reverses its  
 (radial) direction of motion.  

 Let us first consider the case $\Pi^2(r)={\bf p}^2=0$.  The 
 effective potential is as in Ref.~\cite{Kutasov}, where the 
 dynamics of $D$-branes moving in the vicinity of $NS5$-branes 
 has been analysed and explicit solutions have been obtained  
 for $h(r)=1+c/r^2$, $c={\rm const}$.  The motion is restricted 
 to the region: 
 \be\label{constr_Pi_p_zero} 
  h(r)\ge \frac{\mu^2}{{\cal E}^2} \, .  
 \ee 
 Since $h(r\rightarrow\infty)=1$, for $\mu<{\cal E}$ the constraint 
 is empty and the string can escape to infinity.  On the other hand, 
 if $\mu>{\cal E}$ the string does not have enough energy to escape 
 the potential, so it will reach a finite maximum distance, where  
 $V_{\rm eff}=0$, and then reverse its motion to return to $r$=0. 

 Now consider non-zero angular momentum, $\Pi^2(r)>0$, in which case 
 the effective potential is: 
 \be\label{Veff_ang_mom}
  V_{\rm eff} = \frac{g^{rr}}{h(r)} \left( \frac{\mu^2
  +\Pi^2(r)}{h(r){\cal E}^2} - 1 \right) \, . 
 \ee  
 There is now a new possibility in the case ${\cal E}>\mu$: since 
 $\Pi^2(r)$ scales with the inverse metric (see equation 
 (\ref{Pi_of_r})), the angular momentum increases as the string 
 approaches the potential minimum and can dominate over the tension 
 term at short distances.  The effect of the angular momentum is 
 to reduce the radial velocity of the string, until it reaches  
 $V_{\rm eff}=0$, where it bounces and then escapes to infinity.
 For ${\cal E}<\mu$, the string does not have enough energy to 
 escape the attractive potential, and at large enough distances 
 the effective potential approaches a constant value: 
 \be\label{Veff_assympt} 
  V_{\rm eff} \simeq g^{rr}\left( \frac{\mu^2 }{{\cal E}^2}-1 
  \right) \, .
 \ee
 The string reaches a maximum distance and then returns to $r=0$. 
 The angular momentum generally slows down the approach to $r=0$, 
 and solutions include examples where the string spirals an infinite 
 amount of times before reaching the potential centre \cite{Kutasov}.     
 Bound orbits can also exist, depending on the structure of $h(r)$.  
 Indeed, bound orbits were found in the case of KT and KS backgrounds 
 in Ref.~\cite{EaGrTaZa}, where $D3$-brane motion was studied in 
 detail using the Dirac action, and including the relevant Wess-Zumino 
 term.  

 Let us finally focus on the 3D momenta ${\bf p}^2$.  These have 
 the effect of `renormalising' the second term in equation  
 (\ref{Veff_ang_mom}) by a correction of ${\bf p}^2/{\cal E}^2=
 \dot{\bf x}^2 \equiv v_x^2$, that is: 
 \be\label{Veff_p2}
  V_{\rm eff} = \frac{g^{rr}}{h(r)} \left[ \frac{\mu^2
  +\Pi^2(r)}{h(r){\cal E}^2}-\left(1-v_x^2\right) \right] \, .
 \ee 
 In other words, some of the kinetic energy of the string is in 
 the transverse motion in the Minkowskian directions, so it is harder
 for the string to escape to infinity.  At large $r$ the effective 
 potential approaches 
 \be\label{Veff_p2_assympt}
  V_{\rm eff} \simeq g^{rr}\left[ \frac{\mu^2 }{{\cal E}^2} - \left(
  1-v_x^2\right)\right] \, , 
 \ee 
 and so the string needs to have an energy greater than the relativistic 
 mass, ${\cal E}^2>\mu^2/(1-v_x^2)$, in order to escape.  For small 
 distances, where the angular momentum term dominates, the motion 
 changes direction at smaller $r$, as the radial motion of the string 
 is slower than in the $v_x=0$ case.  Note that as $v_x\rightarrow 1$ 
 we must have $\dot r \rightarrow 0$, due to the constraint $1-v^2>0$  
 arising from the square root in the action (\ref{action_sym}).   

 Let us now compare our results with the findings of the previous 
 section.  There is general agreement in the type of orbits that 
 can arise, namely deflections, bounces and bound orbits, but there
 are also some important differences.  In particular, in this section
 the 3D velocities $v_x$ were constant, while in section \ref{VOS} 
 there was a week dependence of $v_x$ on the internal velocity 
 $v_\ell$.  By looking at equation (\ref{v_xdt_w_l_zero}), this 
 dependence can be traced to the string curvature and Hubble 
 friction terms.  While the macroscopic model of the previous 
 section enabled us to allow for correlation-scale curvature 
 giving rise to rms string 3D velocities, here we have considered 
 a straight string with zero curvature.  The most important 
 difference with the previous section is the inclusion of 
 friction terms due to cosmic expansion.  This, for example, 
 changes the dependence of the orbits on $v_x$, as, in the case 
 of no friction, a small $v_x$ generally implies a larger value for 
 the radial velocity in the internal dimensions (as we just saw), 
 but when one includes Hubble friction a small $v_x$ also comes 
 with a stronger damping term on $v_\ell$, which can lead to a 
 smaller internal velocity (Fig.~\ref{vx_dependence}).  Here, 
 there is no Hubble friction in the 10D picture, since the metric 
 is a warped product of Minkowski (as opposed to FLRW in section
 \ref{VOS}) spacetime with a compact internal manifold.  

 To make contact with the previous section, we can move to an 
 effective 4D description by `integrating out' the compact dimensions 
 on the worldsheet action.  Splitting the metric into a 4D and an 
 internal part, $g^{(4)}_{\mu\nu}$ and $g^{(6)}_{\ell m}$ respectively 
 (which include the relevant warping factors), the induced metric on  
 the worldsheet is: 
 \be\label{ind_metric_10D} 
  \gamma^{(10)}_{\alpha\beta}=g^{(4)}_{\mu\nu} \pd_\alpha x^\mu 
  \pd_\beta x^\mu + g^{(6)}_{\ell m} \pd_\alpha l^\ell \pd_\beta
  l^m \, . 
 \ee  
 Defining the induced 4D metric as 
 \be\label{ind_metric_4D}  
  \gamma^{(4)}_{\alpha\beta}=g^{(4)}_{\mu\nu} \pd_\alpha x^\mu 
  \pd_\beta x^\mu \, , 
 \ee
 one can factorise it in the worldsheet Lagrangian, to obtain 
 an effective 4D string action: 
 \bq    
  -\frac{\cal L}{\mu}&=&\sqrt{-{\rm det}\gamma^{(10)}}=\sqrt{ 
  -{\rm det} [\gamma^{(4)}_{\alpha\beta}(\delta^\beta_\gamma + 
  \gamma_{(4)}^{\beta\delta} \, \pd_\delta l^\ell\pd_\gamma l^m 
  \, g^{(6)}_{\ell m} )]} \nonumber \\ 
  &=&\sqrt{-{\rm det}\gamma^{(4)}} 
  \sqrt{ {\rm det}(\delta^\beta_\gamma+\gamma_{(4)}^{\beta\delta}  
  \, \pd_\delta l^\ell\pd_\gamma l^m \, g^{(6)}_{\ell m})} 
  \label{factorise} \, .
 \eq
 Then, using 
 \be\label{expand_matrix} 
  {\rm det}({\bf 1} + {\bf M}) = 1 + \frac{1}{2}{\rm Tr}({\bf M}) 
  - \frac{1}{4}{\rm Tr}({\bf M}^2) + \frac{1}{8}({\rm Tr}{\bf M})^2 
  + {\cal O} ({\bf M}^3) \, ,   
 \ee
 one finds kinetic terms for the worldsheet scalar fields $l^\ell$.  
  
 One can similarly obtain a low-energy Einstein-Hilbert term, starting 
 from the 10D gravitational action.  The 4D metric will then couple 
 to any scalar fields arising from the compactification.  In the 
 setups we are interested in, all scalar fields are stabilised except 
 one, which corresponds to the position of a mobile $D$-brane moving 
 towards an anti-$D$-brane at the bottom of the warping throat.  The 
 interaction between the brane-anti-brane pair, gives rise to a 
 potential for the scalar field, which, subject to fine tuning, can
 satisfy the slow-roll conditions for inflation.  Thus, as the branes 
 approach each other, the scalar field drives inflation in the 
 effective 4D description.  The inflationary phase ends with the 
 collision of the branes and the production of an interacting 
 network of cosmic $D$- and $F$-strings \cite{DvalVil,PolchStab}.  
 The universe enters a radiation-dominated era, so the strings 
 soon find themselves evolving in a power-law FLRW, rather than 
 inflationary, background.   

 We can then understand the essence of our previous results also in 
 this picture: firstly, the string action we found contains kinetic 
 terms for worldsheet scalar fields $l^\ell$, which from the 10D point 
 of view correspond to the positions of the strings in the internal 
 manifold.  These can be thought of as worldsheet currents, which are
 known to result in a reduction of the velocity of strings \cite{book}. 
 We found the same effect from the 10D point view in section \ref{VOS}, 
 were some of the kinetic energy of the string was in the internal 
 directions, so the 
 3D motion of strings was reduced as a result of the constraint 
 $v^2\lesssim 1$ (local) or $v^2\lesssim 1/2$ (for rms velocities 
 in networks) \cite{EDVOS}.  Further, as the strings evolve in an 
 expanding background, there is Hubble friction, which could in 
 principle kill these worldsheet excitations.  In the 10D picture, 
 we found, by considering the string equations of motion, that 
 Hubble damping in the expanding dimensions couples only weakly 
 to the internal excitations 
 and is insufficient to damp them away.  From the low-energy point 
 of view we have just considered, this is still the case because 
 Hubble damping is important on large scales, while the worldsheet 
 scalar excitations operate over much shorter length-scales, over 
 which the background can be taken to be flat.  Further, Hubble 
 damping is becoming less and less important (scaling as $t^{-1}$) 
 on fixed scales, so it can only have a transient effect at early 
 times.  This is in sharp contrast with the case of inflation, 
 where damping can have a much more significant impact 
 (Fig.~\ref{damping}).

 \section{\label{discuss}Discussion} 

 Let us summarise and comment on our results.  In the first part 
 of this paper, we investigated the effect of warping on string 
 evolution, in the case where the background is a warped product 
 of a FLRW universe with a static, toroidal internal space.   
 Starting from the Nambu-Goto equations of motion for strings 
 evolving in this warped background, we identified a number of 
 extra terms that tend to pull the strings towards the bottom of 
 the throat.  We then obtained equations for the velocity evolution  
 of string segments, and, by solving them near the minimum of a  
 warping potential, we quantified the tendency of strings to move 
 towards the minimum.  We noted that, in classical theory, there 
 is not enough damping to guarantee that the strings actually reach 
 the potential minimum and stabilise.  Instead, our analysis supports 
 a picture in which strings oscillate around the bottom, being only  
 weakly damped by cosmological expansion, rather than quickly migrating 
 to it.  During these oscillations, we have found that the 3D string 
 velocity $v_x$ also exhibits oscillatory modulation in its magnitude, 
 due to its coupling to $v_\ell$, but this effect is too small to be 
 of observational significance.  Including angular momentum, and  
 considering different initial conditions, we have found a number 
 of different string trajectories that include deflections, bounces, 
 and bound orbits around the minimum.   

 We then moved on to study string motion in 10D warped backgrounds, 
 like the ones arising in IIB compactifications in brane inflation, 
 where the metric is a warped product of Minkowski spacetime and  
 a Calabi-Yau manifold.  Through a qualitative analysis in terms 
 of an effective potential for one-dimensional radial motion, we 
 found similar string trajectories.  Then, by integrating out the 
 internal dimensions, we obtained kinetic terms of worldsheet 
 scalars, corresponding to the internal string positions in the 
 10D picture.  In the effective 4D picture, one can then understand 
 the effect of slowing down of strings from a slightly different  
 point of view, namely in terms of worldsheet currents as in 
 superconducting strings.  Hubble friction is then inefficient 
 on short scales and decays as $t^{-1}$ during radiation/matter 
 domination.    
   
 In both pictures, our classical analysis points out the absence 
 of strong enough damping to ensure that a generic string trajectory  
 around the potential minimum would be one spiraling towards it, 
 loosing energy on the way, and falling on it.  Instead, we find  
 that generic trajectories near the tip involve series of 
 bounces/deflections with insignificant kinetic energy damping. 
 This could clearly have important implications for string 
 evolution, in particular it could further reduce the average 
 probability of string intercommutations \cite{JoStoTye2,PolchProb}.   
 The effect may not be as dramatic as it looks at first sight, 
 because, in the typical brane inflation setup, the branes collide 
 at the tip of the throat, and so the strings are actually 
 produced close to the bottom.  Thus, if the energy of the 
 produced string is not enough to escape the potential pull, 
 the string can enter a series of bounces, or a bound orbit,  
 its motion being confined within a maximum distance from the  
 centre.  However, depending on the initial internal velocity of 
 the string, this distance may be large enough for it to be a 
 bad approximation to consider the string located at the bottom.  
 It is also possible to produce strings with enough energy to 
 escape the throat region, though this looks statistically unlikely. 
 Indeed, a simple classical estimate obtained by comparing the 
 string potential and kinetic energies (arising from expanding the
 energy (\ref{energy}) in powers of $v_\ell^2\equiv h(r)(g_{rr} 
 \dot r^2 + g_{\theta\phi} \dot l^\theta \dot l^\phi)$ to write
 it as rest mass plus potential and kinetic energy), suggests 
 that the string will escape for $v_\ell^2\gtrsim (h-1)(1-v_x^2)$.  
 Deep in the warping potential where $(h-1)\gg 1$, this is a 
 rare possibility for a scaling network, as there is a `Virial 
 theorem' imposing that $v_x^2+v_\ell^2\simeq 1/2$.  However, 
 since we are now considering strings that were just produced 
 with velocities $v_x, v_\ell$ at the brane collision and had 
 no time to `virialise', this condition does not apply. 
 Unfortunately, the details of the brane collision and annihilation 
 process are at present poorly understood and one cannot quantify 
 the transverse velocity distribution of the produced strings.  
 Presumably, the collision is highly non-adiabatic, giving rise 
 to significant string velocities in the transverse directions, 
 but, due to the Brownian 3D spatial structure of the produced 
 network and local energy conservation, these are expected to be 
 subdominant compared to the corresponding 3D velocities~\cite{EDVOS}.

 It follows, therefore, that the main concern here is not 
 about the strings escaping the warping potential, but, rather,
 entering a series of bounces and deflections about the 
 bottom of the throat, remaining within a maximum distance 
 from it.  This distance defines a volume factor, which 
 suppresses the string intercommuting probability.  Under the 
 assumption that the strings are localised at the bottom of the 
 throat, the corresponding volume factor appearing in the 
 relevant string amplitude is determined by the `thickness' 
 of the string, or better the extend of the wavefunction 
 characterising the fluctuations of the string position around 
 the bottom~\cite{PolchProb,Jackson}, which is of order few 
 string lengths.  Here, the volume factor can be much larger due 
 to the classical motion of the string, the relevant distance 
 scale being much greater than the string scale\footnote{It 
 is, of course, still smaller than the compactification scale.}.  
 Therefore, depending on the details of the brane collision 
 occurring in the final stages of brane inflation (in particular, 
 on the energy transfered to translational degrees of freedom 
 in the internal dimensions), the intercommuting probabilities 
 for cosmic superstrings may be further reduced, with potentially 
 important implications for string network scaling values 
 \cite{JoStoTye2,Sak,intprob}.  Given the uncertainties in the 
 details of the final stages of brane inflation-in particular 
 the collision and annihilation of the branes-it is not possible 
 at present to quantify the expected suppression in the 
 intercommuting probability.  It is clear, however, from the 
 above discussion that this suppression can easily be of one 
 order of magnitude or more.  Thus, combined with recent 
 evidence~\cite{intprob,Vanch_loops} that the scaling string 
 density goes with $P^{-2/3}$ (weaker than initially 
 anticipated), the effects discussed in this paper would
 push up the predicted string densities in these scenarios 
 closer to the initially anticipated levels, obtained  
 by using a larger probability $P$ but stronger dependence 
 of $\rho$ on $P$.
         
 The key point of the no-damping result obtained in this 
 study is that the Nambu-Goto equations of motion imply 
 that Hubble friction in the internal dimensions comes 
 with a velocity-dependent coefficient, which 
 quickly goes to zero as string velocities evolve.  Any 
 other damping term which could operate over cosmological  
 timescales would be enough to ensure string stabilisation  
 at the bottom of the throat, though it seems difficult to 
 motivate such a friction mechanism in classical theory.  One 
 may wonder whether some other mechanism could operate in quantum 
 theory, providing an efficient damping term.  An interesting 
 possibility would be quantum decay to lighter particles, but 
 one would need to know in detail how the worldsheet scalars 
 couple to the standard model and/or other light fields.

\begin{acknowledgments}
The work presented here was initiated by a question of G. Efstathiou.   
I would like to thank Paul Shellard for collaboration during the early 
stages of this project.  This paper has also benefited from discussions 
with Fernando Quevedo, Jose Blanco-Pillado and Ivonne Zavala.  I would 
like to thank Ed Copeland and Anne Davis for their comments and encouragement.
I acknowledge financial support from the Cambridge Newton Trust and the 
EC Marie Curie Research Training Network ENRAGE.  This work is also 
supported in part by MEC, research grant FPA2007-66665. 
\end{acknowledgments}

\bibliographystyle{JHEP.bst}
\bibliography{Warped}

\end{document}